\documentclass[reprint,11pt,superscriptaddress,onecolumn]{revtex4-1}

\usepackage{amsmath,xcolor}
\usepackage{amsfonts,amssymb}
\usepackage{graphics,graphicx}
\usepackage{srcltx}
\usepackage{mathcomp}   
\usepackage{ulem}       
\usepackage[mathscr]{euscript}
\usepackage{mathrsfs}
\usepackage{tabularx,ragged2e,booktabs}
\usepackage{multirow}

\begin{document}

\title{Quantum cryptography without detector vulnerabilities\\
using optically-seeded lasers}

\begin{abstract}
\noindent Security in quantum cryptography~\cite{Gisin,Scarani} is continuously challenged by inventive attacks~\cite{Gerhardt,Lydersen,Xu,Zhao,Qi} targeting the real components of a cryptographic setup, and duly restored by new counter-measures~\cite{Fung09,Yuan,Sas14} to foil them. Due to their high sensitivity and complex design, detectors are the most frequently attacked components. Recently it was shown that two-photon interference~\cite{HOM} from independent light sources can be exploited to avoid the use of detectors at the two ends of the communication channel~\cite{Lo,Braunstein}. This new form of detection-safe quantum cryptography, called Measurement-Device-Independent Quantum Key Distribution (MDI-QKD), has been experimentally demonstrated~\cite{Lo,Tang,Valivarthi,Rubenok,Ferreira,Pirandola15}, but with modest delivered key rates. Here we introduce a novel pulsed laser seeding technique to obtain high-visibility interference from gain-switched lasers and thereby perform quantum cryptography without detector vulnerabilities with unprecedented bit rates, in excess of 1~Mb/s. This represents a 2 to 6 orders of magnitude improvement over existing implementations and for the first time promotes the new scheme as a practical resource for quantum secure communications.
\end{abstract}

\author{L. C. Comandar}
\affiliation{Toshiba Research Europe Ltd, 208 Cambridge Science Park, Cambridge CB4 0GZ, United Kingdom}
\affiliation{Cambridge University Engineering Department, 9 JJ Thomson Ave, Cambridge, CB3 0FA, United Kingdom}

\author{M. Lucamarini}
\email{marco.lucamarini@crl.toshiba.co.uk}
\affiliation{Toshiba Research Europe Ltd, 208 Cambridge Science Park, Cambridge CB4 0GZ, United Kingdom}

\author{B. Fr\"{o}hlich}
\affiliation{Toshiba Research Europe Ltd, 208 Cambridge Science Park, Cambridge CB4 0GZ, United Kingdom}

\author{J. F. Dynes}
\affiliation{Toshiba Research Europe Ltd, 208 Cambridge Science Park, Cambridge CB4 0GZ, United Kingdom}

\author{A.~W.~Sharpe}
\affiliation{Toshiba Research Europe Ltd, 208 Cambridge Science Park, Cambridge CB4 0GZ, United Kingdom}

\author{S. Tam}
\affiliation{Toshiba Research Europe Ltd, 208 Cambridge Science Park, Cambridge CB4 0GZ, United Kingdom}

\author{Z. L. Yuan}
\affiliation{Toshiba Research Europe Ltd, 208 Cambridge Science Park, Cambridge CB4 0GZ, United Kingdom}

\author{R. V. Penty}
\affiliation{Cambridge University Engineering Department, 9 JJ Thomson Ave, Cambridge, CB3 0FA, United Kingdom}

\author{A. J. Shields}
\affiliation{Toshiba Research Europe Ltd, 208 Cambridge Science Park, Cambridge CB4 0GZ, United Kingdom}

\maketitle

\noindent In Quantum Cryptography, a sender Alice transmits encoded quantum signals to a receiver Bob, who measures them and distils a secret string of bits with the sender via public discussion~\cite{Gisin}. Ideally, the use of quantum signals guarantees the information-theoretical security of the communication~\cite{Scarani}. In practice, however, Quantum Cryptography is implemented with real components, which can deviate from the ideal description. This can be exploited to circumvent the quantum protection if the users are unaware of the problem~\cite{blackpaper}.

\begin{figure}
\includegraphics[width=\columnwidth]{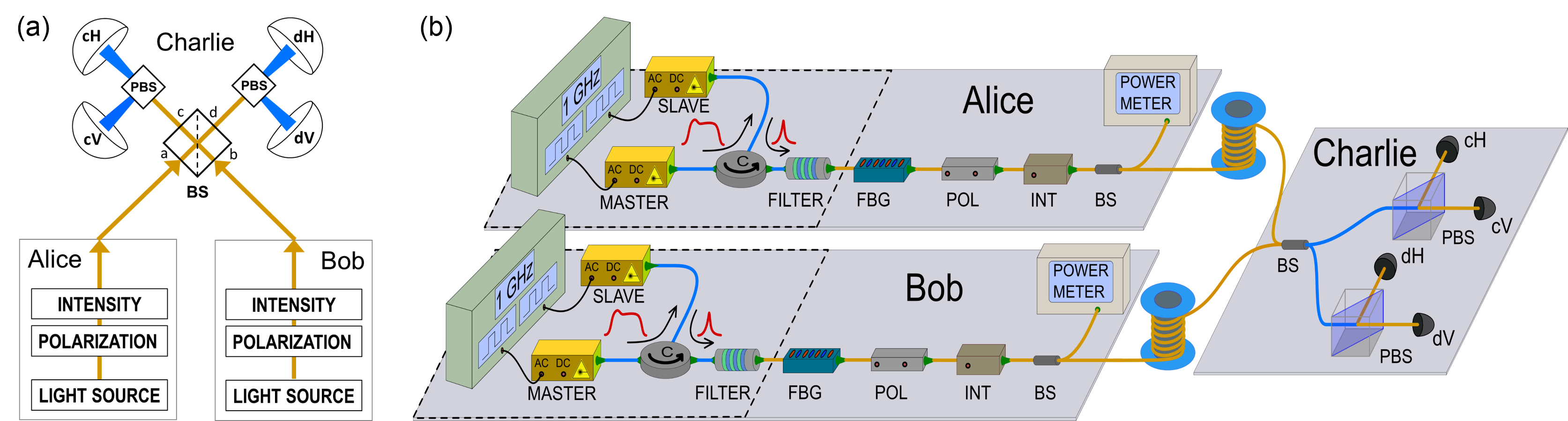}
\caption{
(a) MDI-QKD scheme. Phase-randomised optical pulses are produced by Alice and Bob, set to the desired polarization and intensity, and sent to Charlie. There, they interfere at the beam splitter (BS), pass through the polarizing BS's (PBS's) and reach the four detectors. Coincidence counts from cH/dV or dH/cV (cH/cV or dH/dV) are grouped under the label $|\Psi^-\rangle$ ($|\Psi^+\rangle$), called `singlet' (`triplet')~\cite{Lo}. The measurement outcomes are publicly announced by Charlie.
(b) Experimental MDI-QKD setup. The light sources, which are essential to the results in this work, are enclosed by the dashed lines in Alice's and Bob's setup. C: circulator; FBG: fibre Bragg grating; POL (INT): polarization (intensity) module.
}
\label{Figure1}
\end{figure}

Usually the most complex components are also the most vulnerable. Therefore the vast majority of the attacks performed so far have targeted Bob's single photon detectors~\cite{Gerhardt,Lydersen,Xu,Zhao,Qi}.
MDI-QKD~\cite{Lo,Braunstein} is a recent form of Quantum Cryptography conceived to remove the problem of detector vulnerability. As depicted in Fig.~\ref{Figure1}(a), two light pulses are independently encoded and sent by Alice and Bob to a central node, Charlie. This is similar to a quantum access network configuration~\cite{Bernd}, but in MDI-QKD the central node does not need to be trusted and could even attempt to steal information from Alice and Bob. To follow the MDI-QKD protocol, Charlie must let the two light pulses interfere at the beam splitter inside his station and then measure them. The result can disclose the correlation between the bits encoded by the users, but not their actual values, which therefore remain secret. If Charlie violates the protocol and measures the pulses separately, he can learn the absolute values of the bits, but not their correlation. Therefore he cannot announce the correct correlation to the users, who will then unveil his attempt through public discussion. Irrespective of Charlie's choice, the users' apparatuses no longer need a detector and the detection vulnerability of Quantum Cryptography is removed.

This striking feature of MDI-QKD has fostered intense experimental work and various demonstrations have been provided so far~\cite{Lo,Tang,Valivarthi,Rubenok,Ferreira,Pirandola15}. However, to achieve high-visibility interference at Charlie's beam splitter, the light source in previous experiments was set to emit long pulses at modest clock rates, thus restricting the key rate to less than a hundred bit/s (see Table~\ref{table:1}).

Here we demonstrate a novel high-rate source of indistinguishable pulses from gain-switched laser diodes, ideally suited to MDI-QKD. We use a pair of these sources 
each generating $10^9$ pulses per second, thereby achieving Quantum Cryptography immune to detector attacks at key rates exceeding 1~Mbps for the first time. This is orders of magnitude higher than in previous demonstrations and is comparable to the highest values achieved for conventional Quantum Cryptography~\cite{Comandar}. Furthermore we demonstrate operation for channel loss greater than 20~dB, corresponding to over 100 km of standard fibre. Implementation with real fibre and the effect of a finite sample have also been considered in the experiment.

\begin{table}
\centering
\includegraphics[width=0.75\columnwidth]{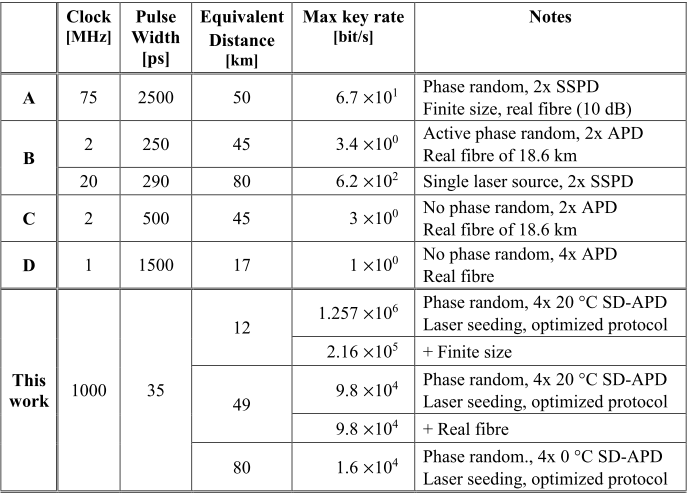}
\caption{Key rates in existing MDI-QKD experiments and comparison with this work. Letters A-D correspond to Refs.~\cite{Tang,Valivarthi,Rubenok,Ferreira}, respectively. Low source clock rate (2nd column) and large pulse width (3rd column) have been used in previous experiments to achieve high-visibility two-photon interference. In one case (B, lower line) a single laser has been employed for both users. Obtaining high visibility from two independent light sources at 1~GHz clock rate and 35~ps pulse width is a major challenge, solved in this work, and can dramatically increase the key rate of MDI-QKD. SSPD: superconducting single photon detector; SD: self-differencing; APD: avalanche photo diode.}\label{table:1}
\end{table}

\begin{figure}
\includegraphics[width=\columnwidth]{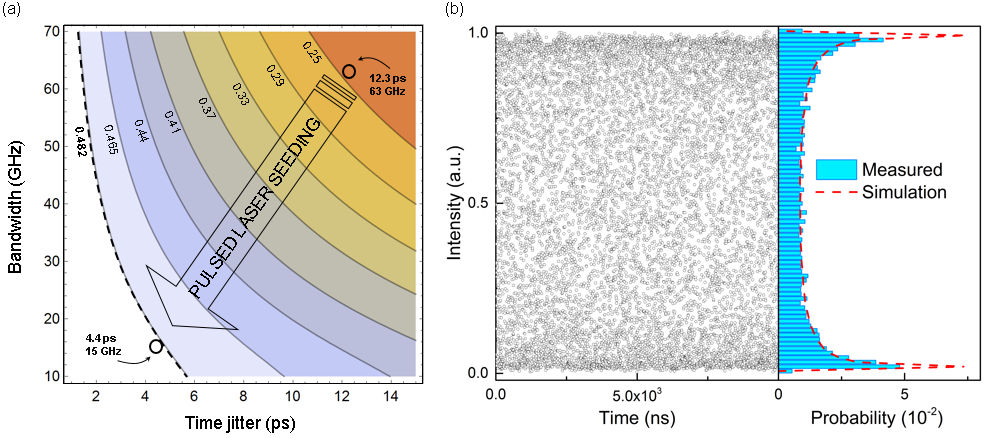}
\caption{
(a) Theoretical contour plot of the two-photon interference visibility versus emission time jitter (horizontal axis) and bandwidth (vertical axis) of the pulses. The arrow shows how pulsed laser seeding improves the measured time jitter and bandwidth (empty circles), thus enhancing the interference visibility. The dashed black line depicts the maximum measured visibility.
(b) Intensity data points and corresponding probability distribution from first-order interference between two consecutive pulses emitted by a seeded laser. The profile of the distribution suggests that the pulses have a random phase~\cite{Yuan2}.
}
\label{Figure3}
\end{figure}

To suit MDI-QKD, the light sources in Alice and Bob have to match stringent criteria. They should emit indistinguishable pulses, to enable high-visibility two-photon interference~\cite{HOM}, and at the same time each pulse should display a random optical phase, to meet a fundamental security condition~\cite{LP07}.
In most demonstrations so far~\cite{Lo,Valivarthi,Rubenok,Ferreira}, light pulses have been carved from a continuous-wave laser. However, the pulses generated this way have a constant or slowly varying optical phase thus violating the random phase condition. An external phase modulator can obviate this problem~\cite{Valivarthi}, but at the expense of increasing cost and complexity of the setup. Semiconductor gain-switched laser diodes can naturally generate short optical pulses ($<$50~ps) with random phases~\cite{Yuan2}. However, the emitted light pulses display a substantial time jitter due to the random nature of the spontaneous emission starting the lasing action. Furthermore they have also a significant spectral width, far exceeding the time-bandwidth limit, due to the frequency chirping arising from transient variation of carrier density in the active medium. These effects combine dramatically to reduce the visibility of the interference. As theoretically depicted in Fig.~\ref{Figure3}(a), temporal jitter and chirp lead to a poor visibility (upper-right corner of the figure).
This has so far prevented the use of gain-switched laser diodes to achieve high-speed MDI-QKD.

Here we propose a novel technique based on pulsed laser seeding to produce low-jitter close-to-transform-limited phase-randomized light pulses from gain-switched lasers. A master laser injects photons into the cavity of a second slave laser through an optical circulator, see Fig.~\ref{Figure1}(b). The lasing action of the slave laser is then initiated by stimulated emission from the light of the master laser rather than by its own spontaneous emission, thus reducing the uncertainty in its emission time. Furthermore, the competition between the cavity modes of the slave laser is immediately resolved by the presence of the master laser's light, thus narrowing the bandwidth of the emitted pulses. The combined effect increases the visibility of the interference between the two narrow pulses emitted by the users' slave lasers. Moreover, the pulsed laser seeding guarantees that the phase of each slave laser is inherited from its own master laser. Due to the fact that the master laser is gain-switched, the master pulse, and hence the slave pulse, has a random optical phase~\cite{Yuan2}.

The improvement in the interference visibility achieved via the pulsed seeding technique is visible from Fig.~\ref{Figure3}(a), where time jitter, bandwidth and visibility of the light sources are experimentally measured and compared against the theoretical prediction. Without pulsed seeding, time jitter and bandwidth of the source amount to 12.3~ps and 63~GHz, respectively, leading to a poor visibility of 25\% and therefore to low key rates. With pulsed laser seeding, on the contrary, they become as small as 4.4~ps and 15~GHz, respectively. For these values we expect an interference visibility of 48.5\%, in good agreement with the experimentally measured value of 48.2\% and close to the theoretical maximum of 50\%~\cite{RTL05}.
The phase randomisation of the pulses emitted by the seeded slave laser is confirmed in Fig.~\ref{Figure3}(b), where the intensity probability distribution has the typical profile expected from the interference of two pulses with random relative phase~\cite{Yuan2}.

\begin{figure}
\includegraphics[width=\columnwidth]{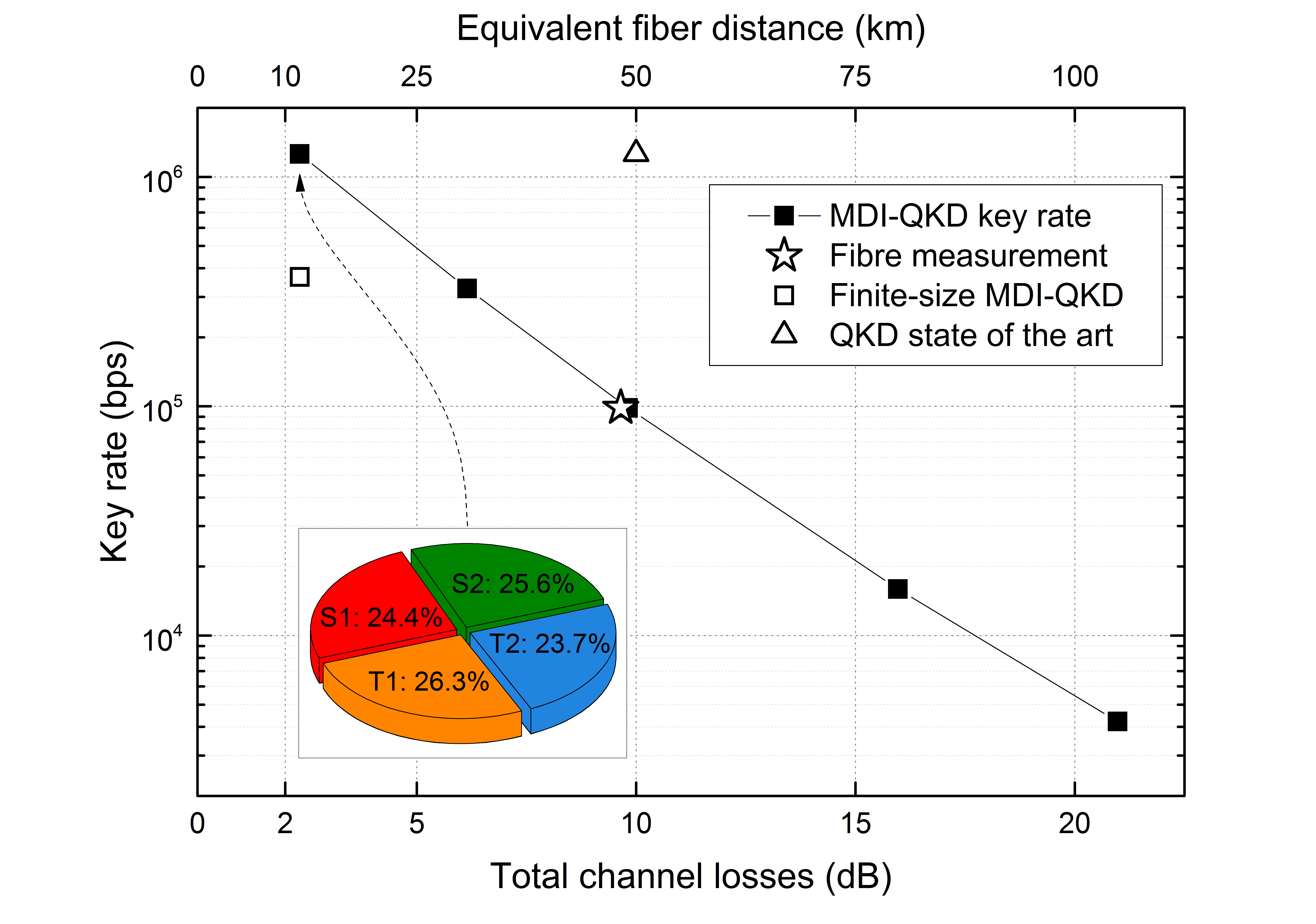}
\caption{MDI-QKD key rates versus total attenuation (lower axis) and equivalent fibre distance (upper axis) of the quantum channel. Solid squares refer to the key rates obtained by varying the channel attenuation in the setup. The empty star is for the rate obtained using two 25-km reels of single mode fibre. The empty square represents the rate after the finite size of the data sample is taken into account. In this case, the total sample size is $\sim 2.4 \times 10^7$, acquired in 12,000 seconds. For comparison, we also add the highest observed finite-size key rate to date for conventional QKD (empty triangle)~\cite{Comandar}. The pie-chart contains the distribution of coincidence counts among the four possible outcomes of Charlie's measurement, for the 2.33~dB loss case. With reference to Fig.~\ref{Figure1}(a), S1, S2, T1 and T2 indicate the coincidence counts of detectors cH/dV, dH/cV, cH/cV and dH/dV, respectively.}
\label{Figure2}
\end{figure}
We performed a series of MDI-QKD experiments using the setup in Fig.~1(b). The results are summarized in Fig.~\ref{Figure2} and detailed in~\cite{SuppInf} (see also Table~\ref{table:1}). The data points represented by the solid squares are obtained by using variable attenuators to reproduce the attenuation of standard single mode fibres ($0.2$~dB/km). The leftmost point corresponds to a rate of $1.257$~Mbit/s, a record in terms of key rate mediated by two-photon interference. The rightmost point corresponds to about 4~kbit/s over 21~dB attenuation, which is still sufficient to generate a 256-AES key more frequently than every 100~ms~\cite{Choi}.

Following the analysis in Ref.~\cite{Ma12}, we consider how the statistical fluctuations of a finite data sample can affect the secure key rate. With the empty square point in Fig.~\ref{Figure2}, we report the finite-size key rate of the system for a 2.33~dB attenuation channel. It amounts to 366~kbit/s and is obtained by gathering the counts from the triplet and singlet states. The finite-size dependence on the channel attenuation is observed to have a similar slope as that in the asymptotic regime.

To replicate a real deployment scenario, we replace the channel attenuation with two single mode fibre spools of 25 km each. We employ two dispersion compensation modules designed for 20~km to cancel the broadening of the pulses due to the chromatic dispersion in the fibre. We also compensate the temporal drift of the arrival time of the pulses at Charlie's beam splitter due to temperature variations. We find that the distilled key rate (empty star in Fig.~\ref{Figure2}) is almost identical to the one obtained from the channel attenuation, proving that fibre-induced effects can be effectively mitigated. All the points in Fig.~\ref{Figure2} have been numerically simulated along the lines of Refs.~\cite{Xu2,Yuan3} to confirm the results and optimize the system.

To illustrate the progress entailed by these results, we report in Fig.~\ref{Figure2} with an empty triangle the state-of-the-art point~\cite{Comandar} of finite-size decoy-state QKD, for a distance of 50 km under similar detection conditions as in the present experiment. Quite impressively, the QKD key rate is only one order of magnitude higher than the corresponding MDI-QKD asymptotic rate, the difference being largely accountable to the non-unitary efficiency of the single photon detectors.

These results prove that MDI-QKD can distribute keys at rates similar to conventional Quantum Cryptography and promote it as a practical solution to serve real-world secure communications.

\section*{Methods}

\subsection*{Experimental setup}

\noindent Alice and Bob consist of two independent pulsed laser seeding-enabled light sources producing phase-randomised 35~ps-long laser pulses at 1550~nm at the repetition rate of 1~GHz. Variable optical band pass filters with 20~GHz bandwidth are aligned to remove any spurious emission. Fibre Bragg gratings are added to pre-compensate for the pulse broadening in the fibre experiment. Polarization and intensity of the pulses are set as required in the protocol and power meters are used to monitor the average photon fluxes. This lets each user prepare weak coherent states in one of four polarization states: $H$, $V$ (rectilinear basis, or $Z$) or $D$, $A$ (diagonal basis, or $X$). The $Z$ basis is used to distil the key bits, while the $X$ basis is used to test the noise on the quantum channel. Alice and Bob select the intensity of the states among four possible values, or ``classes'': $s$ (signal), $u$ (decoy 1), $v$ (decoy 2), $w$ (vacuum). This is different from previous protocols~\cite{Lo}, where three intensity settings rather than four were used, and allows for higher key rates~\cite{Zhou}. For the class $s$, they assign a polarization state from the $Z$ basis, either $H$ or $V$. For the other classes, they assign a polarization state from the $X$ basis, either $D$ or $A$. The intensity is in the range of 0.7~photons/pulse for the $Z$ basis and between 0 and 0.08~photons/pulse for the $X$ basis.
The users send the resulting states to Charlie, with probabilities $p_s=p_Z=1-p_X=45/48$ and $p_u=p_v=p_w=p_X/3$. In Charlie, the beam splitter output ports are spliced to the polarizing beam splitter input ports to ensure polarization alignment to the rectilinear basis and reduce losses. Four InGaAs self-differencing avalanche photodiodes are gated at 1~GHz and synchronised to the arrival time of the photons with 1~ns intrinsic deadtime. Under these conditions the detectors have an effective active window of around 100~ps and are able to measure up to 500~Mcps~\cite{Comandar2}. Their efficiency is kept close to 30\% for the whole duration of the experiment. For attenuation levels up to 16~dB it is advantageous to operate detectors at room temperature (20 $^{\circ}$C), to reduce the afterpulse probability, while for larger attenuation values it is beneficial to operate them at 0 $^{\circ}$C, to produce a smaller dark count rate. At 20~$^{\circ}$C and 0~$^{\circ}$C, the afterpulse probability amounts to 6.5\% and 8.6\%, respectively, and the dark count probability per gate is $6.50\times10^{-5}$ and $2.64\times10^{-5}$, respectively.
Temporal overlap between the pulses is initially achieved by maximizing the single counts within the detection window of the gated detector. This is then fine-tuned by directly measuring the interference visibility in the matched $Z$ basis.

\subsection*{Pulsed laser seeding}

\noindent Each user is endowed with two gain-switched lasers, one of which acts as the master and the other as the slave. The two lasers are driven by square waves at 1~GHz through their AC port. An electrical delay allows to vary the timing between the two driving signals. The DC level of the master laser is set below the threshold ensuring a random phase, but is sufficiently high to have little turn-on delay and produce $\sim$250~ps pulses.  That of the slave is set low enough to assure that no lasing is possible in the absence of the master laser photons.  With seeding photons from the master laser, the slave laser produces pulses around 35~ps wide and close to the time-bandwidth product limit.
In Fig.~\ref{Figure3}(a), the time jitter and frequency bandwidth of the pulses are measured using a fast sampling oscilloscope and an optical spectrum analyser. To test the visibility of the setup we perform a two-photon interference experiment using superconducting single photon detectors. The photon count rate was tuned to give $\sim 10^6$ counts/s per detector. Data was acquired for 50 seconds using a time window of 350~ps around the central peak resulting in a visibility value of 0.482.
For the data presented in Fig.~\ref{Figure3}(b), we use an asymmetric Mach-Zehnder interferometer with an added delay of 1~ns in one arm connected to the output of a seeded slave laser. The interference intensity between subsequent pulses is measured using a PIN photodiode and an oscilloscope. The histogram presented is the result of an acquisition of $10^5$ points.

\section*{Acknowledgments}

\noindent L.~C.~Comandar acknowledges personal support via the EPSRC funded CDT in Photonic Systems Development {and Toshiba Research Europe Ltd}.

\section*{Author contributions}

\noindent Measurements and calculations were performed by L.C.C. and M.L., respectively. The system was readied by B.F., J.F.D., A.W.S., and S.T. Z.L.Y. and A.J.S. conceived the experiment and guided the work. L.C.C. and M.L. wrote the manuscript with contributions from the other authors. All authors discussed experiments, results and the interpretation of results.

%
%

\section*{Supplementary Information}

\subsection{Protocol}
\label{Eprot}

\noindent To increase the final key rate, we adopt an optimised protocol, similar to the one described in~\cite{Zhou5}. It makes use of four intensity settings rather than three to decouple the data basis $Z$ from the test basis $X$. This allows a large photon flux in the $Z$ basis, in the range of 0.7 photons/pulse, thus resulting in a high count rate, in the order of tens of millions counts per second for short distances. It also allows a small photon flux in the $X$ basis, which is optimal for the parameter estimation based on decoy states~\cite{W055,LMC055}. Moreover, we perform decoy state estimation through a numerical routine based on linear programming, detailed below. This increases size and stability of the resulting key rate. All the relevant experimental settings and rates for this protocol are given below in Tables~\ref{tab:I}, \ref{tab:II}, \ref{tab:III}, \ref{tab:IV}.

The steps of the protocol are as follows:

\noindent \underline{Preparation}:
Alice and Bob prepare phase-randomised weak coherent states with mean photon number $\mu_{i}$ (Alice) and $\mu_{j}$ (Bob). The mean photon number $\mu_l$, $l=\{i,j\}$, is randomly chosen among four possible values~\cite{Zhou5}: $s$ (signal), $u$ (decoy~1), $v$ (decoy~2), $w$ (vacuum). When $\mu_l=s$, the users randomly assign a polarization state from the $Z$ basis, either $H$ or $V$. When $\mu_l \neq s$, they randomly assign a polarization state from the $X$ basis, either $D$ or $A$. They send the resulting state to Charlie with probabilities $p_s=p_Z=1-p_X=45/48$ and $p_u=p_v=p_w=p_X/3$.

\noindent \underline{Detection and announcement}:
Charlie performs a Bell measurement on the incoming states. In every run, if at least two detectors click, Charlie publicly announces which detectors clicked. From the announcement, the users draw the successful events, defined as the coincidence counts from two detectors associated to orthogonal ($H/V$) polarizations. When more than two detectors are announced, the users draw all pairwise detector events compatible with the announcement. With reference to Fig.~1, a coincidence count from detectors cH/dV or cV/dH (cH/cV or dH/dV) is assigned to the \textit{singlet} state $\left\vert \psi^{-}\right\rangle=1/\sqrt{2}(\left\vert HV \right\rangle-\left\vert VH \right\rangle)$ (\textit{triplet} state $\left\vert \psi^{+}\right\rangle=1/\sqrt{2}(\left\vert HV \right\rangle+\left\vert VH \right\rangle)$)~\cite{Lo5}.

\noindent \underline{Sifting}:
Alice and Bob announce the bases ($Z$ or $X$) for all the successful events. They perform sifting by keeping the results whenever they have used identical bases and discard the others. Bob performs a bit flip of all his measured bits, except when the matching basis is $X$ and the successful event is a triplet~\cite{Lo5}.

\noindent \underline{Key distillation}:
From the sifted bits, Alice and Bob quantify the gains $Q_{ZZ}^{(k)}$, $Q_{XX}^{(k)}$ and the error rates $E_{ZZ}^{(k)}$, $E_{XX}^{(k)}$ separately for each basis and for the triplet ($k=T$) and singlet ($k=S$) states. From the $X$-basis quantities the users estimate the single photon gain $q_{X}^{(k)}$ and the error rate $e_{X}^{(k)}$ using the decoy state technique. They then use $q_{X}^{(k)}$ to infer a lower bound on $q_{Z}^{(k)}$.
The final key rate $R$ of the system is determined by the rate equations: $R=R^{(T)}+R^{(S)}$; $R^{(k)}=q_{Z}^{(k)}[1-h(e_{X}^{(k)})]-f_{EC}Q_{ZZ}^{(k)}h(E_{ZZ}^{k})$, where $k=\{T,S\}$, $h$ is the binary entropy function and $f_{EC}=1.16$ quantifies how close to the Shannon limit the error correction (EC) performs.

\noindent \underline{Post-processing}: The users run error correction, privacy amplification and all the due post-processing to obtain the final key and secure the overall communication. In this protocol, only results from the rectilinear basis are error-corrected and privacy-amplified. Diagonal basis results are used only for the estimation of the single photon quantities and do not contribute to the raw keys.

\subsection{Distillation procedure}
\label{E}

\noindent In the described protocol, the users distill two separate key rates, one for the singlet ($\left\vert \Psi ^{-}\right\rangle$) and one for the triplet ($\left\vert \Psi ^{+}\right\rangle$) state. The final key rate is given by the sum of the two contributions. Here, we aim to explain the distillation procedure that links the raw count rates to the final key bits. It can be applied to the singlet and triplet data sets separately or to the data set obtained by gathering the data from the two states in a single group. This makes the index $k$ redundant and we drop it from the following discussion. We can then rewrite the key rate equation given above in a more explicit way:
\begin{equation}
R=p_{Z}^{2}\left(  s_{i}e^{-s_{i}}\right)  \left(  s_{j}e^{-s_{j}}\right)
y_{Z}^{1,1}\left[  1-h\left(  e_{X}^{1,1}\right)  \right]  -f_{EC}
Q_{ZZ}^{s_{i},s_{j}}h\left(  E_{ZZ}^{s_{i},s_{j}}\right) .
\label{S2}
\end{equation}
In Eq.~\eqref{S2}, $Q_{ZZ}^{s_{i},s_{j}}$ and $E_{ZZ}^{s_{i},s_{j}}$ are the gain
and the error rate, respectively, measured in the rectilinear basis (see Table~\ref{tab:II}) when
Alice (index `$i$') and Bob (index `$j$') send weak coherent states with mean photon numbers (or ``classes'') $s_{i}$ and $s_{j}$, respectively. We note that according to our protocol, the `$s$' class is selected whenever the basis $Z$ is chosen, so the probability $p_s$ to prepare $s$ coincides with $p_Z$. The key bits are extracted from the $s$ class only, whereas the classes $u$, $v$ and $w$ are used to perform the decoy state estimation~\cite{W055,LMC055}. Because independent weak coherent states are used, the probability that the users simultaneously emit a single photon is the product of two Poisson distributions, which appears in the first pair of brackets in Eq.~\eqref{S2}. The quantity
$y_{Z}^{1,1}$ is the single photon yield in the $Z$ basis, i.e., the
probability that Charlie declares a detection given that Alice and Bob sent
out single photon signals. This quantity is not measurable without true single
photon sources and has to be estimated using the decoy state technique.
Similarly, the quantity $e_{X}^{1,1}$ is the single photon error rate, i.e.,
the probability that the users detect an error from Charlie's declared data
given that they sent out single photon signals, and has to be estimated using
the decoy state technique.

\subsubsection*{Decoy state estimation }
\label{E1}

\noindent We perform the decoy state estimation using a constrained optimization numerical routine similar to Ref.~\cite{T125}.
The first step is to estimate a lower bound for the quantity $y_{Z}^{1,1}$ in Eq.~\eqref{S2}, to lower bound the key rate $R$. This is a
typical constrained minimization problem, where $y_{Z}^{1,1}$ is the objective
function and the constraints are given by the counts acquired in the
experiment. However, instead of minimizing $y_{Z}^{1,1}$, we minimize $y_{X}^{1,1}$, i.e. the single photon yield in the $X$ basis. In the
asymptotic limit, this is justified by the equality $y_{Z}^{1,1}=y_{X}^{1,1}$.
In the finite-size case, we can use the fact that, provided that the sample in the $X$ basis is smaller than the one in the $Z$ basis, the lower bound to $y_{X}^{1,1}$ also represents a lower bound to $y_{Z}^{1,1}$~\cite{T145}. The condition about the sizes of the two samples in the bases $Z$ and $X$ is fulfilled in our experiment, due to the higher photon flux used in the $Z$ basis.

For the objective function $y_{X}^{1,1}$, the constraints
are given by the coincidence counts in the $X$ basis. By dividing the coincidence counts in the $X$ basis (see Table~\ref{tab:III} below) by the number of pulses sent by the users in the $X$
basis (see following section), we can obtain the quantities $Q_{XX}^{\mu_{i},\mu_{j}}$, which can be
plugged in the following equation~\cite{Lo5}:
\[
Q_{XX}^{\mu_{i},\mu_{j}}=\sum_{m=0}^{\infty}\sum_{n=0}^{\infty}\left(
\frac{\mu_{i}^{m}}{m!}e^{-\mu_{i}}\right)  \left(  \frac{\mu_{j}^{n}}
{n!}e^{-\mu_{j}}\right)  y_{X}^{m,n}.
\]
Here, $m$ and $n$ indicate the number of photons emitted by Alice and Bob, respectively. When $\mu_i$ and $\mu_j$ run over $u$, $v$ and $w$, we obtain 9 independent equations,
representing the constraints of the problem. We can rewrite
the constraints as:
\begin{align*}
Q_{XX}^{\mu_{i},\mu_{j}}e^{\mu_{i}}e^{\mu_{j}} &  =\sum_{m,n=0}^{\infty}%
\frac{\mu_{i}^{m}}{m!}\frac{\mu_{j}^{n}}{n!}y_{X}^{m,n}\\
&  =\sum_{m,n=0}^{K}\frac{\mu_{i}^{m}}{m!}\frac{\mu_{j}^{n}}{n!}y_{X}%
^{m,n}+\sum_{m=0}^{K}\sum_{n=K+1}^{\infty}\frac{\mu_{i}^{m}}{m!}\frac{\mu
_{j}^{n}}{n!}y_{X}^{m,n}+\\
&  +\sum_{m=K+1}^{\infty}\sum_{n=0}^{K}\frac{\mu_{i}^{m}}{m!}\frac{\mu_{j}%
^{n}}{n!}y_{X}^{m,n}+\sum_{m,n=K+1}^{\infty}\frac{\mu_{i}^{m}}{m!}\frac
{\mu_{j}^{n}}{n!}y_{X}^{m,n}
\end{align*}
From the above equation, the following bounds can be obtained:
\begin{eqnarray}
\label{cond3p}
\sum_{m,n=0}^{K}\frac{\mu_{i}^{m}}{m!}\frac{\mu_{j}^{n}}{n!}y_{X}^{m,n} &\leq& e^{\mu_{i}}e^{\mu_{j}}Q_{XX}^{\mu_{i},\mu_{j}},\\
\label{cond3m}
\sum_{m,n=0}^{K}\frac{\mu_{i}^{m}}{m!}\frac{\mu_{j}^{n}}{n!}y_{X}^{m,n} &\geq& e^{\mu_{i}}e^{\mu_{j}}\left[  Q_{XX}^{\mu_{i},\mu_{j}}-\left(
1-\frac{\Gamma\left(  1+K,\mu_{i}\right)  }{K!}\frac{\Gamma\left(  1+K,\mu_{j}\right)  }{K!}\right)  \right],
\end{eqnarray}
where $\Gamma\left(  a,b\right)  =\int_{b}^{\infty}t^{a-1}e^{-t}dt$ is the
incomplete gamma function. Eqs.~\eqref{cond3p},~\eqref{cond3m} represent a total of $9+9=18$ constraints. However, as it can be seen from Table~\ref{tab:III}, the counts obtained from the classes $vw$, $wv$ and $ww$ are much fewer than those from the other classes, leading to larger statistical fluctuations. In this case, we found it advantageous to combine the counts from these classes and rewrite the associated constraints in a single cumulative constraint (see also~\cite{YZW5}). This reduces the total number of constraints for the above-described problem to $7+7=14$. Finally, in addition to Eqs.~\eqref{cond3p},~\eqref{cond3m}, we also set the condition that the yields are probabilities, i.e., $y_{X}^{m,n}\in[0,1]$ for every $m, n$.

For the single photon error rate we adopt a procedure similar to the one just described. This time,
we need to maximize the quantity $e_{X}^{1,1}$.
The constraints are given by the following equations~\cite{Lo5}:
\begin{equation*}
Q_{XX}^{\mu_{i},\mu_{j}}E_{XX}^{\mu_{i},\mu_{j}}=e^{-\mu_{i}}e^{-\mu_{j}}\sum_{m=0}^{\infty}\sum_{n=0}^{\infty}\frac{\mu_{i}^{m}}{m!}\frac{\mu_{j}^{n}}{n!}y_{X}^{m,n}e_{X}^{m,n},
\end{equation*}
which leads to the following bound:
\begin{eqnarray}\label{cond5p}
\nonumber Q_{XX}^{\mu_{i},\mu_{j}}E_{XX}^{\mu_{i},\mu_{j}}e^{\mu_{i}}e^{\mu_{j}}
&=&\sum_{m=0}^{\infty}\sum_{n=0}^{\infty}\frac{\mu_{i}^{m}}{m!}\frac{\mu_{j}^{n}}{n!}y_{X}^{m,n}e_{X}^{m,n}\\
\nonumber &=&y_{X}^{0,0}e_{X}^{0,0}+\sum_{n=1}^{\infty}\frac{\mu_{j}^{n}}{n!}y_{X}^{0,n}e_{X}^{0,n}+\sum_{m=1}^{\infty}\frac{\mu_{i}^{m}}{m!}
y_{X}^{m,0}e_{X}^{m,0}+\sum_{m=1}^{\infty}\sum_{n=1}^{\infty}\frac{\mu_{i}^{m}}{m!}\frac{\mu_{j}^{n}}{n!}y_{X}^{m,n}e_{X}^{m,n}\\
\nonumber &=&\frac{1}{2}y_{X}^{0,0}+\frac{1}{2}\sum_{n=1}^{\infty}\frac{\mu_{j}^{n}}{n!}y_{X}^{0,n}+\frac{1}{2}\sum_{m=1}^{\infty}\frac{\mu_{i}^{m}}{m!}
y_{X}^{m,0}+\sum_{m=1}^{\infty}\sum_{n=1}^{\infty}\frac{\mu_{i}^{m}}{m!}\frac{\mu_{j}^{n}}{n!}y_{X}^{m,n}e_{X}^{m,n}\\
&\geq&\frac{1}{2}y_{X}^{0,0}+\frac{1}{2}\sum_{n=1}^{K}\frac{\mu_{j}^{n}}{n!}y_{X}^{0,n}+\frac{1}{2}\sum_{m=1}^{K}\frac{\mu_{i}^{m}}{m!}y_{X}^{m,0}
+\sum_{m=1}^{J}\sum_{n=1}^{J}\frac{\mu_{i}^{m}}{m!}\frac{\mu_{j}^{n}}{n!}y_{X}^{m,n}e_{X}^{m,n}.
\end{eqnarray}
In the third line we set $e_{X}^{0,0}=e_{X}^{0,n}=e_{X}^{m,0}=\frac{1}{2}$ and in
the last line we have dropped some non-negative terms from the sum.
From the last line of Eq.~\eqref{cond5p}, we carry on only the terms corresponding to $J=1$ and drop all the remaining ones. Because the dropped terms are non-negative, we can write:
\begin{equation}
\label{J1}
Q_{XX}^{\mu_{i},\mu_{j}}E_{XX}^{\mu_{i},\mu_{j}}e^{\mu_{i}}e^{\mu_{j}}
\geq\frac{1}{2}y_{X}^{0,0}+\frac{1}{2}\sum_{n=1}^{K}\frac{\mu_{j}^{n}}
{n!}y_{X}^{0,n}+\frac{1}{2}\sum_{m=1}^{K}\frac{\mu_{i}^{m}}{m!}y_{X}^{m,0}
+\mu_{i}\mu_{j}y_{X}^{1,1}e_{X}^{1,1}.
\end{equation}
This can be rewritten as:
\begin{align}
\nonumber e_{X}^{1,1} &  \leq\frac{1}{\mu_{i}\mu_{j}y_{X}^{1,1}}\left(  Q_{XX}^{\mu
_{i},\mu_{j}}E_{XX}^{\mu_{i},\mu_{j}}e^{\mu_{i}}e^{\mu_{j}}-\frac{1}{2}%
y_{X}^{0,0}-\frac{1}{2}\sum_{n=1}^{K}\frac{\mu_{j}^{n}}{n!}y_{X}^{0,n}%
-\frac{1}{2}\sum_{m=1}^{K}\frac{\mu_{i}^{m}}{m!}y_{X}^{m,0}\right)  \\
&  \leq\frac{1}{\mu_{i}\mu_{j}\underline{y_{X}^{1,1}}}\left(  Q_{XX}^{\mu
_{i},\mu_{j}}E_{XX}^{\mu_{i},\mu_{j}}e^{\mu_{i}}e^{\mu_{j}}-\frac{1}{2}%
y_{X}^{0,0}-\frac{1}{2}\sum_{n=1}^{K}\frac{\mu_{j}^{n}}{n!}y_{X}^{0,n}%
-\frac{1}{2}\sum_{m=1}^{K}\frac{\mu_{i}^{m}}{m!}y_{X}^{m,0}\right),
\label{cond5pE}
\end{align}
where in the last line we have indicated with $\underline{y_{X}^{1,1}}$ the lower bound to $y_{X}^{1,1}$ obtained in the previous yield minimization problem.
Eq.~\eqref{cond5pE} represents a set of 9 constraints. As mentioned for the single photon yield optimization problem, the data sets obtained from the classes $vw$, $wv$ and $ww$ are much smaller than the others, so it is beneficial to gather them. We write explicitly the constraints for the least significant classes:
\begin{align*}
e_{X}^{1,1} &  \leq\frac{1}{\underline{y_{X}^{1,1}}}\times\frac{1}{vw}\left(
Q_{XX}^{vw}E_{XX}^{vw}e^{v}e^{w}-\frac{1}{2}y_{X}^{0,0}-\frac{1}{2}\sum
_{m=1}^{K}\frac{v^{m}}{m!}y_{X}^{m,0}-\frac{1}{2}\sum_{n=1}^{K}\frac{w^{n}%
}{n!}y_{X}^{0,n}\right)  \\
e_{X}^{1,1} &  \leq\frac{1}{\underline{y_{X}^{1,1}}}\times\frac{1}{wv}\left(
Q_{XX}^{wv}E_{XX}^{wv}e^{w}e^{v}-\frac{1}{2}y_{X}^{0,0}-\frac{1}{2}\sum
_{m=1}^{K}\frac{w^{m}}{m!}y_{X}^{m,0}-\frac{1}{2}\sum_{n=1}^{K}\frac{v^{n}%
}{n!}y_{X}^{0,n}\right)  \\
e_{X}^{1,1} &  \leq\frac{1}{\underline{y_{X}^{1,1}}}\times\frac{1}{w^{2}%
}\left(  Q_{XX}^{ww}E_{XX}^{ww}e^{2w}-\frac{1}{2}y_{X}^{0,0}-\frac{1}{2}%
\sum_{m=1}^{K}\frac{w^{m}}{m!}y_{X}^{m,0}-\frac{1}{2}\sum_{n=1}^{K}\frac
{w^{n}}{n!}y_{X}^{0,n}\right)
\end{align*}
By adding the three inequalities above we obtain the following cumulative constraint:
\begin{align}
\nonumber e_{X}^{1,1} &  \leq\frac{1}{3\underline{y_{X}^{1,1}}}\left[  \frac{1}{vw}\left(  Q_{XX}^{vw}E_{XX}^{vw}e^{v}e^{w}-\frac{1}{2}\sum_{m=0}^{K}%
\frac{v^{m}}{m!}y_{X}^{m,0}-\frac{1}{2}\sum_{n=1}^{K}\frac{w^{n}}{n!}y_{X}^{0,n}\right)  +\right.  \\
\nonumber  & +\frac{1}{wv}\left(  Q_{XX}^{wv}E_{XX}^{wv}e^{w}e^{v}-\frac{1}{2}\sum_{m=0}^{K}\frac{w^{m}}{m!}y_{X}^{m,0}-\frac{1}{2}\sum_{n=1}^{K}\frac{v^{n}
}{n!}y_{X}^{0,n}\right)  + \\
&  \left.  +\frac{1}{w^{2}}\left(Q_{XX}^{ww}E_{XX}^{ww}e^{2w}-\frac{1}{2}\sum_{m=0}^{K}\frac{w^{m}}{m!}y_{X}^{m,0}-\frac{1}{2}\sum_{n=1}^{K}
\frac{w^{n}}{n!}y_{X}^{0,n}\right)  \right].
\label{condSE}
\end{align}
The $7$ constraints given in Eqs.~\eqref{cond5pE} and \eqref{condSE} have to be added to the 14 specified for the yields in Eqs.~\eqref{cond3p} and~\eqref{cond3m} (including the mentioned cumulative constraint), thus providing a total of 21 constraints for the maximization of $e_{X}^{1,1}$. In addition to these constraints, we also specify in the problem the range of the quantities $y_X^{m,n}$, which is the closed interval $[0,1]$.

\subsubsection*{Finite size key rate}
\label{E2}

\noindent To derive the secure key rate in the presence of the statistical fluctuations of the finite sample, we follow the approach in Ref.~\cite{Ma125} (see also~\cite{Curty5} for a detailed analysis of this subject). We assume that the statistical fluctuations obey a Gaussian distribution~\cite{Ma055}. Therefore it is possible to set the desired failure probability $\varepsilon$ of the estimation procedure by solving the equation $1-\operatorname{erf}\left(  n/\sqrt{2}\right)  =\varepsilon$, where $n$ is the (not necessarily integer) number of standard deviations adding up to form the statistical error of the measured value. We find it convenient to set $n=7$ and obtain $\varepsilon=2.56\times10^{-12}$. Because in our decoy state estimation we use $21$ constraints, this choice assures that the overall failure probability of the estimation of the parameters is less than $5.4\times10^{-11}$.

We then consider the fluctuation function:
\begin{equation}
F(x,n)=\frac{n}{\sqrt{x}},
\label{Ffun}
\end{equation}
where $x$ represents the size of the considered data sample. This function is used to make the constraints in the optimization problems for $y_X^{1,1}$ and $e_X^{1,1}$ looser. For
example, the inequality in Eq.~\eqref{cond3p}
\[
\sum_{m,n=0}^{K}\frac{\mu_{i}^{m}}{m!}\frac{\mu_{j}^{n}}{n!}y_{X}^{m,n}\leq
e^{\mu_{i}}e^{\mu_{j}}Q_{XX}^{\mu_{i},\mu_{j}}
\]
in the finite-size scenario becomes:
\[
\sum_{m,n=0}^{K}\frac{u^{m}}{m!}\frac{v^{n}}{n!}y_{X}^{m,n}\leq e^{u}
e^{v}Q_{XX}^{u,v}\left[  1+F\left(  N_{XX}^{u,v}Q_{XX}^{u,v},7\right)
\right]  ,
\]
where $N_{XX}^{u,v}$ is the total number of runs where Alice and Bob emitted pulses
in the class $u$ and $v$, respectively. Because the resulting constraint is looser, the finite-size solution is always worse than the one in the asymptotic scenario, and the key rate is reduced. This explains why the finite-size key rate for a channel attenuation of 2.33~dB is about $30\%$ of the asymptotic key rate when the total size of the sample is $\sim 2.4 \times 10^7$ (see Table~\ref{tab:III}). The presence of a factor $\sqrt{x}$ in the fluctuation function $F$, Eq.~\eqref{Ffun}, explicitly shows that it is always best to gather the counts from the singlet and triplet data sets in a single group to maximize the finite-size key rate. Because the sizes of the separate triplet and singlet data sets are approximately equal, the size of the total sample is about twice as large as the separate samples. This, according to Eq.~\eqref{Ffun}, entails a factor $\sqrt{2}$ advantage in the key rate if the total sample is used.

The key rate obtained by joining the data sets from the singlet and the triplet states amounts to $366$ kbit/s. The number of prepared pulses are $N_{XX}^{u,u}=N_{XX}^{u,v}=N_{XX}^{u,w}=N_{XX}^{v,u}=N_{XX}^{w,u}=(5 \times 10^2)\times( 4 \times10^9$), acquired in $500$ seconds, and $N_{XX}^{v,v}=N_{XX}^{v,w}=N_{XX}^{w,v}=N_{XX}^{w,w}=(1.25\times10^2)\times(4\times10^9)$, acquired in 125 seconds, where $N_{XX}^{\mu_i,\mu_j}$ is the total number of runs where Alice and Bob simultaneously emitted pulses in the class $\mu_i$ and $\mu_j$, respectively.

\newpage

\subsection{Key rates}
\label{A}

\begin{table}[h!]
\begin{tabular}
[c]{|l|c|}\hline
Channel attenuation/distance &
\begin{tabular}
[c]{c}%
Key rate\\
$\lbrack\text{kbit}/\text{s}]$%
\end{tabular}
\\\hline
\multicolumn{1}{|l|}{2.33 dB (11.65~km)} & $1256.5$\\\hline
\multicolumn{1}{|l|}{2.33 dB (11.65~km)
$\vert$
Finite size} & $366.3$\\\hline
\multicolumn{1}{|l|}{6.15 dB (30.75~km)} & $325.8$\\\hline
\multicolumn{1}{|l|}{9.82 dB (49.10~km)} & $98.2$\\\hline
\multicolumn{1}{|l|}{50~km (9.65~dB)
$\vert$
Real fibre} & $98.4$\\\hline
\multicolumn{1}{|l|}{15.97 dB (79.85~km)} & $15.9$\\\hline
\multicolumn{1}{|l|}{20.98 dB (104.9~km)} & $4.2$\\\hline
\end{tabular}
\caption{Key rate $R$ versus channel attenuation (dB) or equivalent distance (km) in a single mode optical fibre featuring 0.2~dB/km attenuation.}\label{tab:I}
\end{table}

\subsection{Count and error rates in the rectilinear basis}
\label{B}

\begin{table}[h!]
\begin{tabular}{|c|c|c|c|c||c|}
\cline{1-6}
Channel attenuation/distance & \multicolumn{2}{c|}{$
\begin{array}{c}
\text{Singlet} \\
\left\vert \Psi ^{-}\right\rangle
\end{array}%
$} & \multicolumn{2}{|c||}{$
\begin{array}{c}
\text{Triplet} \\
\left\vert \Psi ^{+}\right\rangle
\end{array}%
$} & $s_{A}=s_{B}$ \\ \cline{2-6}
& $C_{ZZ}^{(S)}$ & $E_{ZZ}^{(S)}$ & $C_{ZZ}^{(T)}$ & $E_{ZZ}^{(T)}$ & ph/pulse \\ \hline
\multicolumn{1}{|l|}{2.33 dB (11.65~km)} & $288399$ & $0.33\%$ & $287902$ & $%
0.35\%$ & $0.7$ \\ \hline
\multicolumn{1}{|l|}{2.33 dB (11.65~km) | Finite size} & $%
307574$ & $0.29\%$ & $308259$ & $0.25\%$ & $0.7$ \\ \hline
\multicolumn{1}{|l|}{6.15 dB (30.75~km)} & $139817$ & $0.47\%$ & $139345$ & $%
0.52\%$ & $0.7$ \\ \hline
\multicolumn{1}{|l|}{9.82 dB (49.10~km)} & $39881$ & $0.61\%$ & $39993$ & $%
0.63\%$ & $0.7$ \\ \hline
\multicolumn{1}{|l|}{50~km (9.65~dB) | Real fibre} & $53411$
& $0.81\%$ & $54058$ & $0.86\%$ & $0.7$ \\ \hline
\multicolumn{1}{|l|}{15.97 dB (79.85~km)} & $14704$ & $0.95\%$ & $14657$ & $%
1.14\%$ & $0.6$ \\ \hline
\multicolumn{1}{|l|}{20.98 dB (104.9~km)} & $3238$ & $1.20\%$ & $3211$ & $%
0.97\%$ & $0.55$ \\ \hline
\end{tabular}
\caption{Measured coincidence counts ($C_{ZZ}$) and error rates ($E_{ZZ}$) in the rectilinear basis, separately for the singlet and the triplet states. Acquisition time is 80~ms for each attenuation/distance value.}\label{tab:II}
\end{table}

\newpage

\subsection{Count rates in the diagonal basis}
\label{C}

\begin{table}[h!]
\begin{tabular}{|cc|c|c|c|c|c|c||c|}
\cline{1-9}
Channel attenuation/distance&  & \multicolumn{3}{c|}{$%
\begin{array}{c}
\text{Singlet} \\
\left\vert \Psi ^{-}\right\rangle
\end{array}%
$} & \multicolumn{3}{c||}{$%
\begin{array}{c}
\text{Triplet} \\
\left\vert \Psi ^{+}\right\rangle
\end{array}%
$} & $\mu_i=\mu_j$ \\ \cline{2-8}
& \multicolumn{1}{|c|}{Class} & $u$ & $v$ & $w$ & $u$ & $v$ & $w$ & ph/pulse
\\ \hline
\multicolumn{1}{|l}{2.33 dB (11.65~km)} & \multicolumn{1}{|c|}{%
\begin{tabular}{c} $u$ \\$v$ \\$w$%
\end{tabular}%
} &
\begin{tabular}{c}$223041$ \\$84703$ \\$71263$%
\end{tabular}
&
\begin{tabular}{c}$92393$ \\$14119$ \\$9218$%
\end{tabular}
&
\begin{tabular}{c}$80410$ \\$9514$ \\$5516$%
\end{tabular}
&
\begin{tabular}{c}$225777$ \\$85996$ \\$72727$%
\end{tabular}
&
\begin{tabular}{c}$93290$ \\$14362$ \\$9353$%
\end{tabular}
&
\begin{tabular}{c}$80759$ \\$9780$ \\$5778$%
\end{tabular}
&
\begin{tabular}{c}$0.01$ \\$0.002$ \\$0.001$%
\end{tabular}
\\ \hline
\multicolumn{1}{|l}{2.33 dB (11.65~km) | Finite size} &\multicolumn{1}{|c|}{%
\begin{tabular}{c}$u$ \\$v$ \\$w$%
\end{tabular}%
} &
\begin{tabular}{c}$4771407$ \\$1774040$ \\$1506023$%
\end{tabular}
&
\begin{tabular}{c}$1967827$ \\$73783$ \\$46317$%
\end{tabular}
&
\begin{tabular}{c}$1693057$ \\$47749$ \\$27113$%
\end{tabular}
&
\begin{tabular}{c}$4853012$ \\$1773149$ \\$1510257$%
\end{tabular}
&
\begin{tabular}{c}$1969107$ \\$73789$ \\$46591$%
\end{tabular}
&
\begin{tabular}{c}$1690185$ \\$48136$ \\$27348$%
\end{tabular}
&
\begin{tabular}{c}$0.01$ \\$0.002$ \\$0.001$%
\end{tabular}
\\ \hline
\multicolumn{1}{|l}{6.15 dB (30.75~km)} & \multicolumn{1}{|c|}{%
\begin{tabular}{c}$u$ \\$v$ \\$w$%
\end{tabular}%
} &
\begin{tabular}{c}$218848$ \\$93154$ \\$80250$%
\end{tabular}&
\begin{tabular}{c}$79813$ \\$13655$ \\$9458$%
\end{tabular}&
\begin{tabular}{c}$67269$ \\$8681$ \\$5276$%
\end{tabular}&
\begin{tabular}{c}$222021$ \\$93588$ \\$81479$%
\end{tabular}&
\begin{tabular}{c}$80688$ \\$13955$ \\$9410$%
\end{tabular}&
\begin{tabular}{c}$67866$ \\$8710$ \\$5377$%
\end{tabular}&
\begin{tabular}{c}$0.016$ \\$0.0032$ \\$0.0016$%
\end{tabular}
\\ \hline
\multicolumn{1}{|l}{9.82 dB (49.10~km)} & \multicolumn{1}{|c|}{%
\begin{tabular}{c}$u$ \\$v$ \\$w$%
\end{tabular}%
} &
\begin{tabular}{c}$170331$ \\$67118$ \\$57570$%
\end{tabular}&
\begin{tabular}{c}$69123$ \\$11298$ \\$7397$%
\end{tabular}&
\begin{tabular}{c}$59436$ \\$7674$ \\$4694$%
\end{tabular}&
\begin{tabular}{c}$174234$ \\$68020$ \\$57787$%
\end{tabular}&
\begin{tabular}{c}$70675$ \\$11749$ \\$7526$%
\end{tabular}&
\begin{tabular}{c}$60709$ \\$7831$ \\$4759$%
\end{tabular}&
\begin{tabular}{c}$0.025$ \\$0.005$ \\$0.0025$%
\end{tabular}
\\ \hline
\multicolumn{1}{|l}{50~km (9.65~dB) | Real fibre} &\multicolumn{1}{|c|}{%
\begin{tabular}{c}$u$ \\$v$ \\$w$%
\end{tabular}%
} &
\begin{tabular}{c}$239876$ \\$88586$ \\$74187$%
\end{tabular}&
\begin{tabular}{c}$98512$ \\$14526$ \\$9310$%
\end{tabular}&
\begin{tabular}{c}$85722$ \\$9854$ \\$5604$%
\end{tabular}&
\begin{tabular}{c}$244648$ \\$89817$ \\$74644$%
\end{tabular}&
\begin{tabular}{c}$98477$ \\$14946$ \\$9275$%
\end{tabular}&
\begin{tabular}{c}$85958$ \\$9721$ \\$5745$%
\end{tabular}&
\begin{tabular}{c}$0.025$ \\$0.005$ \\$0.0025$%
\end{tabular}
\\ \hline
\multicolumn{1}{|l}{15.97 dB (79.85~km)} & \multicolumn{1}{|c|}{%
\begin{tabular}{c}$u$ \\$v$ \\$w$%
\end{tabular}%
} &
\begin{tabular}{c}$226108$ \\$88114$ \\$74616$%
\end{tabular}&
\begin{tabular}{c}$82071$ \\$11477$ \\$7051$%
\end{tabular}&
\begin{tabular}{c}$69065$ \\$6971$ \\$3401$%
\end{tabular}&
\begin{tabular}{c}$225133$ \\$87615$ \\$75312$%
\end{tabular}&
\begin{tabular}{c}$83685$ \\$11730$ \\$7078$%
\end{tabular}&
\begin{tabular}{c}$70665$ \\$7033$ \\$3461$%
\end{tabular}&
\begin{tabular}{c}$0.05$ \\$0.01$ \\$0.005$%
\end{tabular}
\\ \hline
\multicolumn{1}{|l}{20.98 dB (104.9~km)} & \multicolumn{1}{|c|}{%
\begin{tabular}{c}$u$ \\$v$ \\$w$%
\end{tabular}%
} &
\begin{tabular}{c}$142656$ \\$56455$ \\$48101$%
\end{tabular}&
\begin{tabular}{c}$50799$ \\$7057$ \\$4391$%
\end{tabular}&
\begin{tabular}{c}$42995$ \\$4233$ \\$2237$%
\end{tabular}&
\begin{tabular}{c}$145584$ \\$57492$ \\$49104$%
\end{tabular}&
\begin{tabular}{c}$52744$ \\$7308$ \\$4578$%
\end{tabular}&
\begin{tabular}{c}$44096$ \\$4322$ \\$2211$%
\end{tabular}&
\begin{tabular}{c}$0.08$ \\$0.0155$ \\$0.008$%
\end{tabular}\\ \hline
\end{tabular}
\caption{Measured coincidence counts in the diagonal basis ($C_{XX}$) separately for the singlet and the triplet states. Acquisition time is 25 seconds per combination, with the exception of the line containing the finite-size data. There the total sample was acquired in 12,000 seconds.}\label{tab:III}
\end{table}

\newpage

\subsection{Error rates in the diagonal basis}
\label{D}

\begin{table}[h!]
\begin{tabular}{|cc|c|c|c|c|c|c||c|}
\cline{1-9}
Channel attenuation/distance&  & \multicolumn{3}{c|}{$%
\begin{array}{c}
\text{Singlet} \\
\left\vert \Psi ^{-}\right\rangle
\end{array}%
$} & \multicolumn{3}{|c||}{$%
\begin{array}{c}
\text{Triplet} \\
\left\vert \Psi ^{+}\right\rangle
\end{array}%
$} & $\mu_i=\mu_j$ \\ \cline{2-8}
& \multicolumn{1}{|c|}{Class} & $u$ & $v$ & $w$ & $u$ & $v$ & $w$ & ph/pulse
\\ \hline
\multicolumn{1}{|l}{2.33 dB (11.65~km)} & \multicolumn{1}{|c|}{%
\begin{tabular}{c}$u$ \\$v$ \\$w$%
\end{tabular}%
} &
\begin{tabular}{c}$31.82\%$ \\$40.17\%$ \\$44.07\%$%
\end{tabular}&
\begin{tabular}{c}$41.60\%$ \\$38.39\%$ \\$41.53\%$%
\end{tabular}&
\begin{tabular}{c}$45.37\%$ \\$42.21\%$ \\$43.22\%$%
\end{tabular}&
\begin{tabular}{c}$31.55\%$ \\$40.67\%$ \\$44.55\%$%
\end{tabular}&
\begin{tabular}{c}$40.77\%$ \\$37.68\%$ \\$41.28\%$%
\end{tabular}&
\begin{tabular}{c}$44.36\%$ \\$41.95\%$ \\$42.13\%$%
\end{tabular}&
\begin{tabular}{c}$0.01$ \\$0.002$ \\$0.001$%
\end{tabular}
\\ \hline
\multicolumn{1}{|l}{2.33 dB (11.65~km) | Finite size} &\multicolumn{1}{|c|}{%
\begin{tabular}{c}$u$ \\$v$ \\$w$%
\end{tabular}%
} &
\begin{tabular}{c}$33.12\%$ \\$40.95\%$ \\$45.79\%$%
\end{tabular}&
\begin{tabular}{c}$41.20\%$ \\$38.98\%$ \\$41.30\%$%
\end{tabular}&
\begin{tabular}{c}$44.90\%$ \\$41.64\%$ \\$42.61\%$%
\end{tabular}&
\begin{tabular}{c}$32.08\%$ \\$40.56\%$ \\$43.93\%$%
\end{tabular}&
\begin{tabular}{c}$41.62\%$ \\$38.98\%$ \\$41.63\%$%
\end{tabular}&
\begin{tabular}{c}$45.33\%$ \\$41.49\%$ \\$43.39\%$%
\end{tabular}&
\begin{tabular}{c}$0.01$ \\$0.002$ \\$0.001$%
\end{tabular}
\\ \hline
\multicolumn{1}{|l}{6.15 dB (30.75~km)} & \multicolumn{1}{|c|}{%
\begin{tabular}{c}$u$ \\$v$ \\$w$%
\end{tabular}%
} &
\begin{tabular}{c}$32.81\%$ \\$41.56\%$ \\$45.06\%$%
\end{tabular}&
\begin{tabular}{c}$40.06\%$ \\$38.22\%$ \\$42.17\%$%
\end{tabular}&
\begin{tabular}{c}$44.66\%$ \\$41.25\%$ \\$43.33\%$%
\end{tabular}&
\begin{tabular}{c}$30.83\%$ \\$41.06\%$ \\$45.01\%$%
\end{tabular}&
\begin{tabular}{c}$39.92\%$ \\$37.46\%$ \\$39.90\%$%
\end{tabular}&
\begin{tabular}{c}$43.54\%$ \\$41.00\%$ \\$41.27\%$%
\end{tabular}&
\begin{tabular}{c}$0.016$ \\$0.0032$ \\$0.0016$%
\end{tabular}
\\ \hline
\multicolumn{1}{|l}{9.82 dB (49.10~km)} & \multicolumn{1}{|c|}{%
\begin{tabular}{c}$u$ \\$v$ \\$w$%
\end{tabular}%
} &
\begin{tabular}{c}$32.69\%$ \\$41.41\%$ \\$45.17\%$%
\end{tabular}&
\begin{tabular}{c}$41.94\%$ \\$40.23\%$ \\$43.75\%$%
\end{tabular}&
\begin{tabular}{c}$45.88\%$ \\$42.39\%$ \\$44.91\%$%
\end{tabular}&
\begin{tabular}{c}$32.42\%$ \\$40.92\%$ \\$44.90\%$%
\end{tabular}&
\begin{tabular}{c}$41.32\%$ \\$40.07\%$ \\$40.95\%$%
\end{tabular}&
\begin{tabular}{c}$44.08\%$ \\$41.21\%$ \\$43.12\%$%
\end{tabular}&
\begin{tabular}{c}$0.025$ \\$0.005$ \\$0.0025$%
\end{tabular}
\\ \hline
\multicolumn{1}{|l}{50~km (9.65~dB) | Real fibre} &\multicolumn{1}{|c|}{%
\begin{tabular}{c}$u$ \\$v$ \\$w$%
\end{tabular}%
} &
\begin{tabular}{c}$32.72\%$ \\$39.98\%$ \\$43.21\%$%
\end{tabular}&
\begin{tabular}{c}$42.87\%$ \\$39.60\%$ \\$40.49\%$%
\end{tabular}&
\begin{tabular}{c}$46.10\%$ \\$43.00\%$ \\$42.95\%$%
\end{tabular}&
\begin{tabular}{c}$31.45\%$ \\$40.96\%$ \\$45.29\%$%
\end{tabular}&
\begin{tabular}{c}$41.25\%$ \\$37.68\%$ \\$42.24\%$%
\end{tabular}&
\begin{tabular}{c}$44.68\%$ \\$42.41\%$ \\$43.19\%$%
\end{tabular}&
\begin{tabular}{c}$0.025$ \\$0.005$ \\$0.0025$%
\end{tabular}
\\ \hline
\multicolumn{1}{|l}{15.97 dB (79.85~km)} & \multicolumn{1}{|c|}{%
\begin{tabular}{c}$u$ \\$v$ \\$w$%
\end{tabular}%
} &
\begin{tabular}{c}$30.39\%$ \\$39.30\%$ \\$43.31\%$%
\end{tabular}&
\begin{tabular}{c}$39.77\%$ \\$34.94\%$ \\$38.15\%$%
\end{tabular}&
\begin{tabular}{c}$44.48\%$ \\$37.56\%$ \\$37.31\%$%
\end{tabular}&
\begin{tabular}{c}$31.71\%$ \\$41.35\%$ \\$45.25\%$%
\end{tabular}&
\begin{tabular}{c}$38.75\%$ \\$34.53\%$ \\$36.54\%$%
\end{tabular}&
\begin{tabular}{c}$43.13\%$ \\$36.93\%$ \\$39.12\%$%
\end{tabular}&
\begin{tabular}{c}$0.05$ \\$0.01$ \\$0.005$%
\end{tabular}
\\ \hline
\multicolumn{1}{|l}{20.98 dB (104.9~km)} & \multicolumn{1}{|c|}{%
\begin{tabular}{c}$u$ \\$v$ \\$w$%
\end{tabular}%
} &
\begin{tabular}{c}$32.09\%$ \\$41.70\%$ \\$45.32\%$%
\end{tabular}&
\begin{tabular}{c}$40.09\%$ \\$35.47\%$ \\$37.94\%$%
\end{tabular}&
\begin{tabular}{c}$44.18\%$ \\$37.04\%$ \\$37.33\%$%
\end{tabular}&
\begin{tabular}{c}$30.83\%$ \\$39.42\%$ \\$43.24\%$%
\end{tabular}&
\begin{tabular}{c}$39.69\%$ \\$34.21\%$ \\$37.27\%$%
\end{tabular}&
\begin{tabular}{c}$44.12\%$ \\$37.58\%$ \\$36.27\%$%
\end{tabular}&
\begin{tabular}{c}$0.08$ \\$0.0155$ \\$0.008$%
\end{tabular}
\\ \hline
\end{tabular}
\caption{Measured error rates ($E_{XX}$) in the diagonal basis separately for the singlet and the triplet states. Acquisition times are as in Table~\ref{tab:III}.}\label{tab:IV}
\end{table}

\newpage

\subsection{Theoretical estimation of the visibility}
\label{F}

\noindent The two-photon interference visibility $V\left(  \sigma_{\tau},\Delta v\right)$ obtained from two independent gain-switched laser diodes is a function of the time jitter $\tau$ and of the bandwidth $\Delta \nu$ of the interfering pulses, which are reported on the horizontal and vertical axis of Fig.~2(a) in the main text, respectively. Their values for the slave lasers in our setup, with and without the pulse laser seeding technique, have been measured and are given as abscissas and ordinates, respectively, of the two empty circles in the figure. The time jitter is assumed to follow a Normal distribution $N_{\tau}\left(  0,\sigma_{\tau}\right)$ centred at $0$. The visibility is plotted from the expression:
\[
\widetilde{V}\left(  \sigma_{\tau},\Delta v\right)  =\int_{-\infty}^{\infty}d\tau V\left(
\tau,\Delta v\right)  N_{\tau}\left(  0,\sigma_{\tau}\right),
\]
where
\begin{equation}\label{V1}
V\left(  \tau,\Delta v\right)  =\frac{1}{2}\exp\left[  -\frac{\tau
^{2}+4\left(  \omega_{ij}+2\tau\beta\right)^{2}\sigma_{t}^{4}}{4\sigma
_{t}^{2}}\right]  .
\end{equation}
The quantity $V\left(  \tau,\Delta v\right)$ in Eq.~\eqref{V1} is obtained from the electric fields
\begin{equation}
\nonumber
\xi_{l}(t)=\sqrt{I(t)}e^{i\left(  \omega_{l}t+\beta t^{2}+\theta_{l}\right)  }
\end{equation}
emitted by Alice ($l=i$) and Bob ($l=j$), which are used to estimate the coincidence counts at Charlie's detectors~\cite{Xu25,Yuan35}.
The random variable $\tau$ in Eq.~\eqref{V1} represents the total time jitter between the two
pulses emitted by the users measured from Charlie's beam splitter; $\omega_{ij}=2\pi\left(\nu_{j}-\nu_{i}\right)  $ accounts for the (small) difference in the
central frequencies $\nu_{i}$ and $\nu_{j}$ of the interfering pulses emitted by
Alice and Bob, respectively; $\sigma_{t}$ is the standard deviation of the
optical pulses having intensity profile $I(t)$, assumed be Gaussian,
and are related to the measurable full-width-at-half-maximum of the pulses, $\Delta t$, by the
relation $\sigma_{t}=\Delta t/\left(  2\sqrt{2\ln2}\right)  $. The parameter
$\beta$ accounts for the frequency chirp. We assume that frequency chirp is
the only cause of a bandwidth larger than the one prescribed by the
time-bandwidth product. In this case it can be shown that $\Delta
v=\Delta v^{(0)}\sqrt{1+16\beta^{2}\sigma_{t}^{4}}$~\cite{Agrawal}. By measuring
$\sigma_{t}$ and the time-bandwidth product of the emitted pulses, it is possible to invert this
relation and determine the parameter $\beta$.


\end{document}